# Trigger excitation of "pearls" in the Earth's magnetosphere: On the 90th anniversary of the discovery of Pc1 waves

A.V. Guglielmi[1], B.V. Dovbnya[2]

[1]*Institute of Physics of the Earth RAS, Moscow, Russia*,

[2] *Borok Geophysical Observatory of Institute of Physics of the Earth RAS, Borok, Yaroslavl region, Russia*

**Abstract**

Ultra-low-frequency electromagnetic waves Pc1 (0.2 – 5 Hz), widely known in the literature as "pearls", are excited in the outer radiation belt and propagate to the Earth's surface along the geomagnetic field lines in the form of Alfvén waves. The study of pearls is of considerable interest for magnetospheric physics. Both spontaneous and stimulated pearl excitation are observed. The paper examines stimulated (trigger) excitation of pearls. A classification of triggers acting on dynamic systems of the magnetosphere is presented. 2 types, 4 classes and 8 species of triggers have been introduced. Examples of triggers of natural and artificial origin are given. The concept of a trigger cascade is introduced. Particular attention is paid to the anthropogenic periodic trigger of pearls. It manifests itself in the form of the so-called Big Ben effect. The essence of the effect is that a series of pearls is often excited immediately following the hour marker according to world time. It is claimed that the connection between the excitation of pearls and hour markers is not accidental, but represents a rather mysterious geophysical phenomenon. It is assumed that the Big Ben effect occurs as a result of the impact on the radiation belt of an unknown type of endogenous periodic artificial trigger.

## 1. Introduction

We will call a trigger a small disturbance of a dynamic system that leads to a change in its structure and/or functioning. The dynamic system that we will consider in this paper is an ion cyclotron resonator (ICR) that exists in the equatorial zone of the Earth's outer radiation belt [1]. In the ICR, ultra-low-frequency electromagnetic waves Pc1 (0.2 – 5 Hz) are excited, widely known in the literature as "pearls" [2].

The pearls were discovered by Eyvin Sucksdorff at Sodankylä Observatory [3] and Leiv Harang at Tromsø Observatory [4]. This event had a noticeable influence on the development of magnetospheric physics (see, for example, reviews [5–7] and monographs [8, 9], as well as a special issue of the journal J. Atmosph. Solar-Terrestr. Phys. [10], dedicated to the 70th anniversary of the discovery of the pearls). Our work is dedicated to the 90th anniversary of the discovery of pearls. The general view on the origin of pearls is that they are excited in the outer radiation belt and propagate towards the earth's surface along the geomagnetic field lines in the form of Alfvén waves [2, 7].

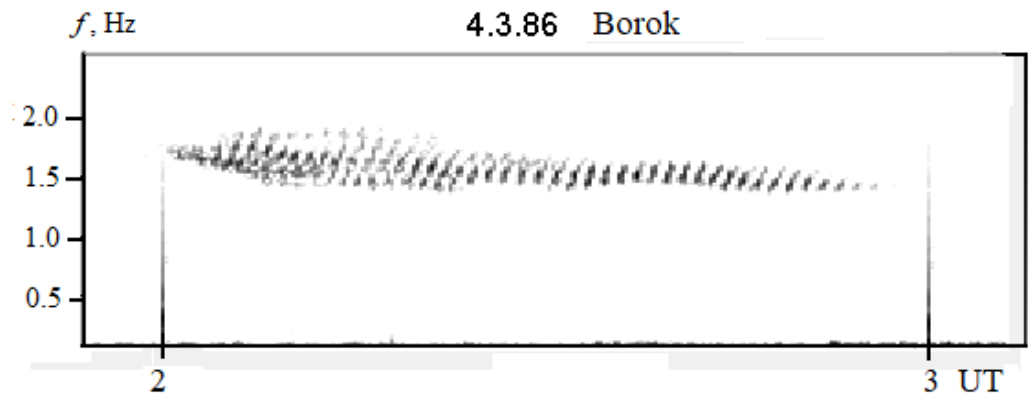

**Fig. 1**. Dynamic spectrum of pearls registered in the Borok Observatory IPE RAS on March 4, 1986.

Figure 1 shows a typical series of pearls. Sometimes a series occurs spontaneously, without any apparent reason, and at times the beginning of a series is preceded by one or another disturbance, which can be considered as a trigger that sets in motion the mechanism for generating pearls in the ICR. We will focus on the trigger excitation of pearls. But first we will present a general classification of triggers acting on the dynamic systems of the magnetosphere.

## 2. Classification of triggers

Let's divide all triggers by origin into two types: natural and artificial (anthropogenic) triggers. Each type is divided into two classes – endogenous and exogenous triggers, depending on whether the trigger occurs inside or outside the magnetosphere. We will divide the classes into species based on the trigger's dependence on time. Let us distinguish between periodic and aperiodic triggers (see table).

**Table**. Classification of triggers

| TRIGGER ||
|---|---|
| **Type** ||
| Natural | Artificial |
| **Class** ||
| Endogenous | Exogenous |
| **Species** ||
| Periodic | Aperiodic |

So, we have 2 types, 4 classes and 8 species of triggers. One could continue the classification by dividing species into varieties and so on, but we will not do this, and will only note that it is reasonable to provide each species with an additional explanation indicating the phenomenon that induces a given trigger. For example, the expression "injection of energetic charged particles into closed shells of the magnetosphere" can serve as such an explanation for a trigger that is a reconnection of geomagnetic lines in the neutral layer of the magnetotail.

It is convenient to assign a three-digit digital designation to each species. Let's assign the number 1 to each category on the left side of the table, and the number 2 to each category on the right side. Then, for example, a change in the sign of the X-component at the boundary between sectors of the interplanetary magnetic field is a trigger of species 121. Let's give one more example to show the classification system in action. The re-alignment of the geomagnetic field lines in the magnetosphere tail mentioned above is a trigger of species 112.

## 3. Natural triggers

In the middle of the last century, the former Minister of Education of the RSFSR, Aleksey Georgievich Kalashnikov, due to life circumstances unrelated to the subject of our discussion, began privately studying geomagnetic pulsations in the range of 0.2 – 5 Hz. Being a physicist, he probably did not for a moment take into account the Epicurean view of the nature of things and immediately focused his attention on finding the cause of the spontaneous activation of pulsations. He put forward the idea that the cause of the activation is meteor showers [11, 12]. The idea was supported [13–17], but was not developed.

The search for a causal relationship between Pc1 and sudden onset of magnetic storms SSC turned out to be more effective [6l. It is highly likely that following the SSC, a series of pearls will be excited in the near-noon sector of the magnetosphere. The SSC trigger should be classified as s0ecies 122.

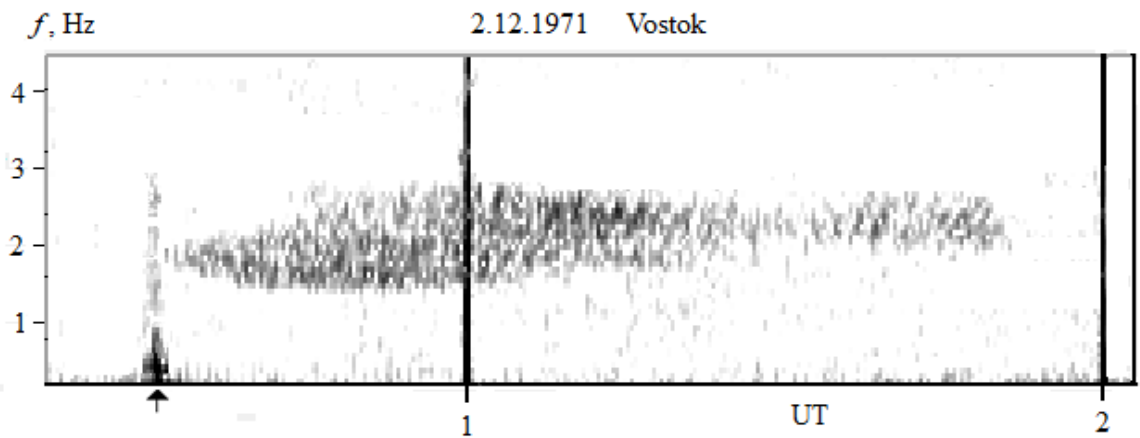

**Fig. 2**. Excitation of a series of pearls following the Pi1B pulse (trigger type 112). The event was observed at Vostok Observatory on December 2, 1971. The short broadband Pi1B pulse is indicated by the arrow.

Let us take a closer look at another natural trigger, namely Pi1B, which belongs to 112 s0ecies. The result of its action is shown in Figure 2. Pi1B geomagnetic pulsations have the form of a short broadband pulse. It is assumed that Pi1B arise as a result of the injection of a fluctuating flow of energetic electrons into the high-latitude layers of the ionosphere in the vicinity of the midnight meridian [5].

It is not known for certain how the Pi1B trigger affects the ICR, switching it into self-excitation mode. Further targeted studies are needed to clarify the relationship between Pi1B properties and the properties of Pc1 induced by this trigger. A favorable circumstance is that the properties of Pi1B themselves have been studied in considerable detail.

## 4. Artificial triggers

Each species of artificial trigger is naturally divided into two varieties: targeted and targetless triggers. Targeted ones arise when conducting active experiments in near-Earth space, for example, producing a controlled powerful explosion, or heating the ionosphere with powerful radio waves [18]. The second variety arises as a by-product of human technological activity. We will present here one side effect of this kind, described for the first time in [19]. In literature it is known as the clock mark effect, or Big Ben effect [8]. Its distinctive feature is as follows. We are confident that we are dealing with an artificial exogenous periodic trigger of specjies 221, since the division of world time into hourly intervals occurred by conditional agreement between countries. The hour mark itself on the sonogram, the striking of the clock on the Spasskaya Tower, or on the Tower of London, of course, has nothing to do with magnetospheric phenomena. The Big Ben effect only indicates the existence of an unknown trigger 221. We don't know what this trigger is as a physical object. Could a radio time signal be a trigger? It is difficult to imagine this, although one should not lose sight of the fact that a huge number of radio pulses affect the near-Earth environment globally, and with high precision, synchronously.

This situation is not unique. Species 221 also includes unknown triggers responsible for the calendar effect of weekends in the activity of geophysical processes [20, 21]. The uncertainty of the physical nature of triggers, on the one hand, challenges our ability to understand natural phenomena, and on the other hand, opens up space for hypothetical assumptions. But we will return to the Big Ben effect in the activity of pearls and try to demonstrate the reality of its existence.

Let's look at Figure 1 again. If we look closely, we will see that it is a magnificent illustration of the Big Ben effect. The series of pearls is excited immediately after the hour mark, shown in the figure by the vertical line. The first time this connection was observed at Tiksi Observatory in 1977 [19]. And such cases are by no means isolated. Let us give a number of illustrative examples.

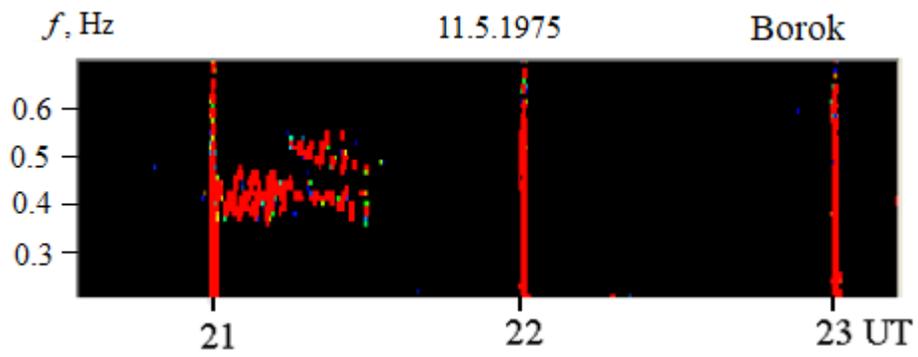

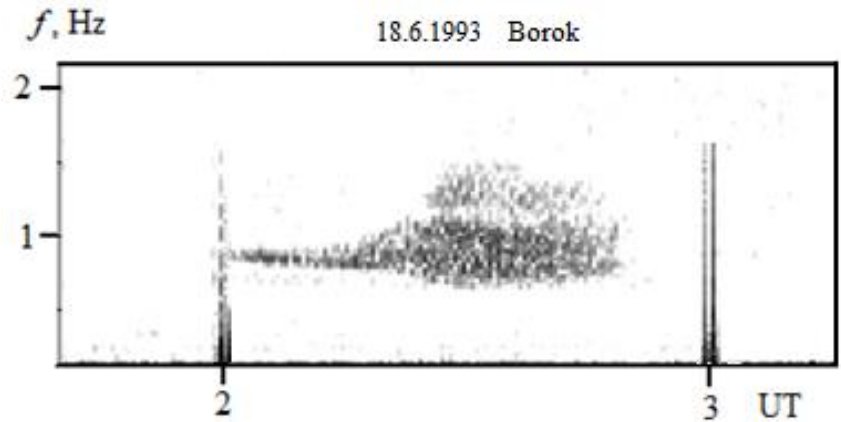

**Fig. 3**. Two clear examples of the Big Ben effect from observations at Borok Observatory on May 11, 1975 and June 18, 1993.

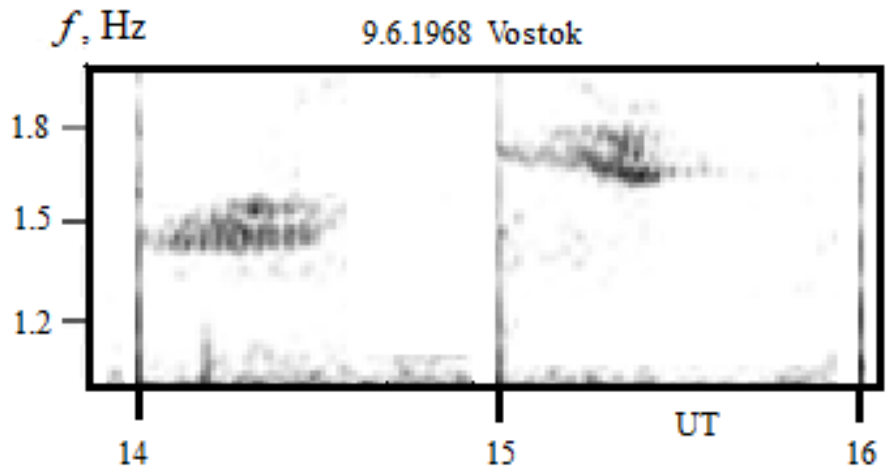

**Fig. 4**. Doublet of anthropogenic triggers of pearls as observed at Vostok Observatory on June 9, 1968.

One of the authors (B.V.) analyzed data from the Borok, Vostok, College, Mondy, and Tiksi observatories for the presence of the Big Ben effect and found that the effect was registered at each of them. The paired Big Ben effect shown in Figure 4 is interesting. Let us pay attention to the amazing similarity in the structure of the two half-hour series of pearls.

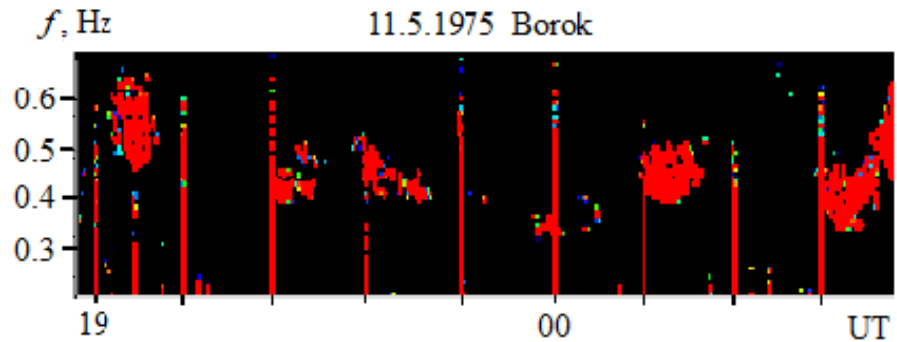

**Fig. 5**. Train of anthropogenic triggers of pearls as observed at the Borok observatory on May 11, 1975.

In our opinion, looking at Figure 5, which shows a fivefold manifestation of the Big Ben effect in eight hours, we must reject the assumption of a simple coincidence of the onset of excitation of the pearls with the hour markers. Further information on the Big Ben effect can be found in articles [22–25].

## 5. Discussion

The question of the mechanism of trigger excitation of pearls is extremely interesting from a physical point of view. It can be discussed in detail in the case of triggers of natural origin, as well as in the case of artificial triggers created in targeted experiments. Knowing the properties of the trigger, we put forward one or another theoretical hypothesis and have the opportunity to test it in an experiment using the classical "input-output" scheme.

However, we will begin the discussion with the question not about the trigger, but about the spontaneous excitation of pearls that appear as if without any apparent reason. It is obvious that the parameters of the ICR change over time. In the linear approximation, the evolution of the oscillation amplitude in the ICR is described by the equation

$$\frac{dA}{dt} = \gamma(\Lambda)A. \tag{1}$$

Инкремент $\gamma$ зависит от управляющего параметра $\Lambda = NT_\perp$, который равен плотности поперечной энергии протонов внешнего радиационного пояса.

The increment $\gamma$ depends on the control parameter $\Lambda = NT_\perp$, which is equal to the transverse energy density of protons of the outer radiation belt. Here $N$ is the concentration of protons, $T_\perp$ is their effective transverse temperature. From the theory of cyclotron instability it follows that

$$\gamma(\Lambda) \propto (\Lambda - \Lambda_c), \tag{2}$$

where $\Lambda_c(T_\parallel, Q)$ is the critical value of the control parameter, $T_\parallel$ is the longitudinal temperature, and $Q$ is the quality factor of the resonator [8]. Let $\Lambda < \Lambda_c$ at first. Oscillations in the ICR are not excited, there are no pearls. If the control parameter $\Lambda$ or the quality factor $Q$ increases monotonically, then sooner or later, it is quite possible, a situation will arise where $\Lambda > \Lambda_c$. Self-excitation of the ICR will occur and a series of pearls will emerge [2].

Now let us discuss the issue of forced excitation of pearls by the SSC. In this case, a rapid compression of the magnetosphere occurs and the mechanism of betatron acceleration of protons of the outer radiation belt is activated. The value of $\Lambda$ increases sharply, and if before SSC there was $\Lambda < \Lambda_c$ and there were no pearls, then after SSC it is quite possible that there will be $\Lambda > \Lambda_c$ and the excitation of a series of pearls will begin.

The situation is more complicated with the Pi1B impulse. The point is this. Along with energetic electrons, protons are also injected into the magnetosphere. It is not entirely clear whether the trigger for the pearls is the intrusion of energetic protons into the midnight sector of the magnetosphere from the geomagnetic tail? We will leave the question unanswered, but it is quite amenable to further study.

Let us note that a cascade of triggers may arise in the magnetosphere. Let us describe the scenario of three-stage trigger excitation of pearls. The first trigger of type 122 is the reconnection of the interplanetary magnetic field lines and the geomagnetic field lines on the frontal surface of the magnetosphere. It induces the excitation of geomagnetic pulsations (see for example [26]). The second stage of the cascade occurs in the magnetotail (trigger 112) in the form of a reconnection of geomagnetic field lines in the neutral layer of the tail. Reconnection is facilitated by Alfvén waves excited by the first trigger. They accelerate the polar wind and saturate it with heavy ions [27]. The jet in the polar wind flow, accelerated and weighted, reaches the neutral layer of the tail and induces reconnection, which gives rise to the formation of the third trigger – the injection of energetic charged particles into closed magnetic shells in the midnight sector of the magnetosphere.

The most difficult thing to understand is the origin of the Big Ben effect. The trigger exists, but is unknown to us, and we cannot analyze the ICR response to anthropogenic disturbance using the "input-output" scheme. The information available to us is contained only in the output signal. It can be assumed that we are dealing with an electromagnetic, rather than a mechanical, version of the 221 trigger. In fact, the technosphere and the outer radiation belt are separated by a huge distance, which cannot be overcome by mechanical disturbance in one or two minutes. Overall, the question of the origin of the Big Ben effect remains open. It deserves further study, since the observed effect indirectly indicates a parasitic pulsation of global energy consumption, synchronized with the world time clock.

We would not like to end the discussion without offering at least a palliative solution to the problem. Let us pay attention to the fact that the radio pulse, polarizing the plasma, excites waves whose phase velocity is less than the group velocity of the radio pulse, similar to how a fast-moving charge in a medium with a refractive index greater than unity excites Vavilov-Cherenkov radiation. Our hypothesis is that a sufficiently powerful radio pulse, excited in the technosphere, excites an Alfvén wave in the ionosphere, which propagates upward and, upon reaching the ICR, switches the resonator into self-excitation mode.

## 6. Conclusion

The pearls discovered 90 years ago by Sucksdorff and Harang provide us with a excellent example of the stable wave structures of the Earth's magnetosphere. Sometimes pearls are excited spontaneously, without any apparent reason, but at times we observe cases of triggered excitement. Experimental and theoretical study of the relationship between the trigger and the subsequent series of pearls is interesting and productive from a physical point of view. The Big Ben effect, discovered by us in 1977, seems especially interesting and mysterious. The trigger exists and is certainly anthropogenic, but we still don't know about it. This complicates the experimental study of the Big Ben effect and poses a complex problem for the theory of excitation and propagation of ultra-low-frequency electromagnetic waves. The existence of the Big Ben effect is not in doubt for us, but firstly, independent confirmation of the effect is absolutely necessary. Secondly, if such confirmation is successful, it is necessary to find an anthropogenic trigger that induces the pearls' excitation.

*Acknowledgments*. We express our sincere gratitude to F.Z. Feygin, A.S. Potapov and O.D. Zotov for their interest in this work and fruitful discussions. The work was carried out within the framework of the planned tasks of the Ministry of Science and Higher Education of the Russian Federation to the Institute of Physics of the Earth of the Russian Academy of Sciences.

This paper will be submitted to the journal "Dynamic Processes in Geospheres".